\documentclass[usenatbib]{mnras}
\usepackage{newtxtext,newtxmath}
\usepackage[T1]{fontenc}
\DeclareRobustCommand{\VAN}[3]{#2}
\let\VANthebibliography\thebibliography
\def\thebibliography{\DeclareRobustCommand{\VAN}[3]{##3}\VANthebibliography}

\usepackage{graphicx}	
\usepackage{amsmath}	
\usepackage{orcidlink}

\newcommand{\Sec}[1]{Section~\ref{#1}}

\usepackage{colortbl}%
\newcommand{\myrowcolour}{\rowcolor[gray]{0.925}}
\newcommand{\cosmorate}{{\sc Cosmo}$\mathcal{R}${\sc ate}}
\newcommand{\mobse}{{\sc mobse{}}}
\usepackage{soul}

\title[BBH spins: model selection with GWTC-3]{Binary black hole spins: model selection with GWTC-3}

\author[C.~P\'{e}rigois et al.]{
Carole~P\'{e}rigois$^{1,2,}$\thanks{E-mail: caroleperigois@outlook.com (CP)}$^{,\orcidlink{0000-0002-9779-2838}}$,
Michela~Mapelli$^{1,2,3,}$\thanks{E-mail:michela.mapelli@unipd.it}$^{,\orcidlink{0000-0001-8799-2548}}$,
Filippo~Santoliquido$^{1,2,3,\orcidlink{0000-0003-3752-1400}}$,
 Yann~Bouffanais$^{1,2,\orcidlink{0000-0003-3462-0366}}$,
\newauthor{~and Roberta~Rufolo$^1$}.
\\
$^{1}$Physics and Astronomy Department Galileo Galilei, University of Padova, Vicolo dell’Osservatorio 3, I–35122, Padova, Italy\\
$^{2}$INFN - Padova, Via Marzolo 8, I–35131 Padova, Italy\\
$^{3}$INAF - Osservatorio Astronomico di Padova, Vicolo dell’Osservatorio 5, I-35122 Padova, Italy
}

\date{Accepted XXX. Received YYY; in original form ZZZ}

\pubyear{2015}

\begin{document}
\label{firstpage}
\pagerange{\pageref{firstpage}--\pageref{lastpage}}
\maketitle

\begin{abstract}
The origin of the spins of stellar-mass black holes is still controversial, and angular momentum transport inside massive stars is one of the main sources of uncertainty. 
Here, we apply hierarchical Bayesian inference to derive constraints on spin models from the 59 most confident binary black hole merger events in the third gravitational-wave transient catalogue (GWTC-3). We consider up to five 
parameters:  chirp mass, mass ratio, redshift, effective spin, and precessing spin. For model selection, we use a  set of binary population synthesis simulations spanning drastically different assumptions for black hole spins and natal kicks. In particular, our spin models  range from maximal to minimal efficiency of angular momentum transport in stars. 
 We find that,  if we include the precessing spin parameter into our analysis, models predicting only vanishingly small spins are in tension with GWTC-3 data. On the other hand, models in which most spins are vanishingly small, but that also include a  sub-population of tidally spun-up black holes are a good match to the data. Our results show that the precessing spin parameter has a crucial impact on model selection. 

\end{abstract}

\begin{keywords}
black hole physics -- gravitational waves --  binaries: general -- stars: black holes
\end{keywords}



\section{Introduction}
\label{sec:intro}

The third observing run (O3) of the Advanced LIGO \citep{adLIGO} and Virgo \citep{adVirgo} detectors has brought the number of compact binary merger observations up to 90 events with 
a probability of astrophysical origin 
$>0.5$ \citep{gwtc1,gwtc2, gwtc21, gwtc3_catalogue}. In particular, the 63 confident detections of binary black hole (BBH) mergers (with a false alarm rate FAR$<0.25$~yr$^{-1}$)  lead to more accurate constraints on the mass and spin distribution of these systems  \citep{gwtc3_population}.


The intrinsic distribution of primary black hole (BH) masses inferred by the LIGO--Virgo--KAGRA collaboration (hereafter, LVK) shows several sub-structures, including a main peak at $\approx{10}$ M$_\odot$, a secondary peak at $\approx{30-40}$~M$_\odot$, 
and a long tail extending up to $\sim{80}$ M$_\odot$ \citep[e.g.,][]{gwtc3_population}. The inferred distribution of mass ratios has a strong preference for equal-mass systems, but several BBHs are confidently unequal-mass \citep[e.g.,GW190412][]{GW190412}\citep[,GW190517][]{gwtc2}. Focusing on BH spins, we can safely exclude that all BHs are maximally spinning \citep{Farr_2017,Farr_2018, gwtc1}. Typical spin magnitudes in BBHs are small, with $\sim{50}$\% of BHs  having 
$\chi \lesssim{} 0.3$ \citep[e.g.,][]{Wysocki_2019,gwtc2}, although not all BHs  in the LVK sample have zero spin \citep{Roulet_2019,Miller_2020}. For example, GW151226 \citep{abbott2016GW151226} and GW190517 \citep{gwtc3_population} confidently possess spin. LVK data also support some mild evidence for spin-orbit misalignment \citep[e.g.,][]{Tiwari_2018,gwtc2,gwtc3_population,Venumadhav_2020,Olsen_2022, Callister_2021, Hannam_2021, Callister_2022}. 

These results provide crucial insights to understand BBH formation and evolution  \citep[e.g.,][]{Gerosa_2013,Stevenson_2015,Rodriguez_2016,Stevenson_2017,Talbot_2017,Fishbach_2017,Vitale_2017,Zevin_2017,Farr_2018,Barrett_2018,Taylor_2018,Fragione_2019,Sedda_2019,Roulet_2019,Wysocki_2019,Bouffanais_2019,Bouffanais_2021a,Bouffanais_2021b,Kimball_2020a,Kimball_2020b,Baibhav_2020,Sedda_2020,Zevin_2021,Mapelli_2021,Mapelli_2022}. Moreover, the mass and spin of BHs carry the memory of their progenitor stars and therefore are a key to unravel the details of massive star evolution and collapse \citep[e.g.,][]{Fryer_2001,Heger_2003,Belczynski_2010,Mapelli_2013,Fragos_2015,Marchant_2016,Eldridge_2016,Demink_2016,Spera_2017,Bavera_2020,Belczynski_2020,Fragione_2021b, Mandel_2021,Fryer_2022,Olejak_2022,Chattopadhyay_2022,Vanson_2022,Briel_2022,Stevenson_2022, Broekgaarden_2022,Broekgaarden_2022b}. In particular, the spin magnitude of a stellar-origin BH should retain the imprint of the spin of the core of its progenitor star \citep[e.g.,][]{Qin_2018,Qin_2019,Fuller_2019,Bavera_2020,Belczynski_2020,Olejak_2021,Stevenson_2022b}. 


Several models have been proposed to infer the spin magnitude of the BH from that of the progenitor star. The main open question concerns the efficiency of angular momentum transport within a star \citep[e.g.,][]{Maeder_2000,Cantiello_2014,Fuller_2019b}. If angular momentum is efficiently transferred from the core to the outer layers, mass loss by stellar winds can dissipate most of it, leading to a low-spinning stellar core and then to a low-spinning BH. If instead the core retains most of its initial angular momentum until the final collapse, the  BH will be fast spinning.

In the shellular model \citep{Zahn_1992,Ekstroem_2012,Limongi_2018,Costa_2019}, angular momentum is mainly transported by meridional currents and shear
instabilities, leading to relatively inefficient spin dissipation. In contrast, according to the Tayler-Spruit dynamo mechanism \citep{Spruit_2002}, differential rotation induces the formation of an unstable magnetic field configuration, leading to an efficient transport of angular momentum via magnetic torques. 
Building upon the Tayler-Spruit mechanism, \cite{Fuller_2019} derived a new model with an even more efficient angular momentum dissipation, predicting that the core of a single massive star might end its life with almost no rotation.

Electromagnetic observations yield controversial results. Asteroseismology favours slowly rotating cores in the late evolutionary stages, but the vast majority of stars with an asteroseismic estimate of the spin are low-mass stars \citep{Mosser_2012,Gehan_2018,Aerts_2019}. 
Continuum-fitting derived spins of BHs in high-mass X-ray binaries are extremely high \citep[e.g.,][]{Reynolds_2021,Miller-jones_2021,Fishbach_2022}, but such  measurements might be affected by substantial observational biases \citep[e.g.,][]{Reynolds_2021}.  Finally, BH spins inferred from quasi periodic oscillations yield notably smaller values than continuum fitting. For example, the estimate of the  dimensionless spin of the BH in GRO~J1655--40 is  $\chi=0.7\pm{}0.1$ and $0.290\pm{} 0.003$ from continuum fitting \citep{Shafee_2006} and quasi-periodic oscillations \citep{Motta_2014}, respectively.


In a binary system, the evolution of the spin is further affected by tidal forces and accretion, which tend to spin up a massive star, whereas non-conservative mass transfer and common-envelope ejection enhance mass loss, leading to more efficient spin dissipation \citep{Kushnir_2016,Hotokezaka_Piran_2017,Zaldarriaga_2018,Qin_2018}. For example, the model by \cite{Bavera_2020} shows that the second-born BH can be highly spinning if its progenitor was tidally spin up when it was a Wolf-Rayet star orbiting about the first-born BH.

Furthermore, the orientation of the BH spin with respect to the orbital angular momentum of the binary system encodes information about binary evolution processes. 
In a tight binary system, tides and mass transfer tend to align the stellar spins with the orbital angular momentum (\citealt{Gerosa_2018}, but see \citealt{Stegmann_2021} for a possible spin flip process induced by mass transfer). If the binary system is in the field, the supernova kick is the main mechanism that can misalign the spin of a compact object with respect to the orbital angular momentum, by tilting the orbital plane \citep[e.g.,][]{Kalogera_2000}. Finally, the spins of BHs in dynamically formed binary systems are expected to be isotropically distributed, because close encounters in a dense stellar cluster reset any previous signature of alignment \citep[e.g.,][]{Rodriguez_2016,Mapelli_2021}.




Here, we perform a model-selection hierarchical Bayesian analysis on confident LVK BBHs 
($p_{\rm astro}\,>\,0.9$ and $\mathrm{FAR}\,<\,0.25\,{\rm yr}^{-1}$). We consider models of field BBHs 
for three of the most used angular-momentum  transport models: (i) the shellular model as implemented in the Geneva stellar evolution code \citep{Ekstroem_2012}, (ii) the Tayler-Spruit dynamo model as implemented in the {\sc mesa} code \citep{Cantiello_2014}, and (iii) the model by \cite{Fuller_2019}.  Hereafter, we will refer to these three models simply as GENEVA (G), MESA (M) and FULLER (F) models, following the description in \cite{Belczynski_2020}.


For each of these models,  we consider an additional variation accounting for the Wolf-Rayet (WR) star tidal spin-up mechanism described by \cite{Bavera_2020}. Also, we account for spin tilts induced by core-collapse supernova explosions.

This paper is organized as follows. \Sec{sec:models} presents our population-synthesis  models. \Sec{sec:analysis} describes the hierarchical Bayesian framework we used and discusses the LVK  events used in our study. We lay down the results  in \Sec{sec:results}, and summarize our conclusions in \Sec{sec:conclusion}.

\section{Astrophysical Models}\label{sec:models}

\subsection{{\sc mobse} and natal kicks}
\label{sec:mobse}

We simulated our binary systems with the code {\sc mobse}  \citep{Mapelli_2017,Giacobbo_2018}. {\sc mobse} is a custom and upgraded version of {\sc bse } \citep{Hurley_2000,Hurley_2002}, in which we introduced metallicity-dependent stellar winds for OB \citep{Vink_2001}, WR \citep{Belczynski_2010}, and luminous blue variable stars \citep{GiacobboMapelli_2018}. {\sc mobse} includes a formalism for electron-capture \citep{GiacobboMapelli_2019}, core-collapse \citep{Fryer_2012}, and (pulsational) pair-instability supernovae \citep{Mapelli_2020}. Here, we adopt the rapid core-collapse supernova prescription, which enforces a gap between the maximum mass of neutron stars and the minimum mass of BHs (2--5 M$_\odot$, \citealt{Oezel_2010,Farr_2011}).

\begin{figure}
    \includegraphics[width=\columnwidth]{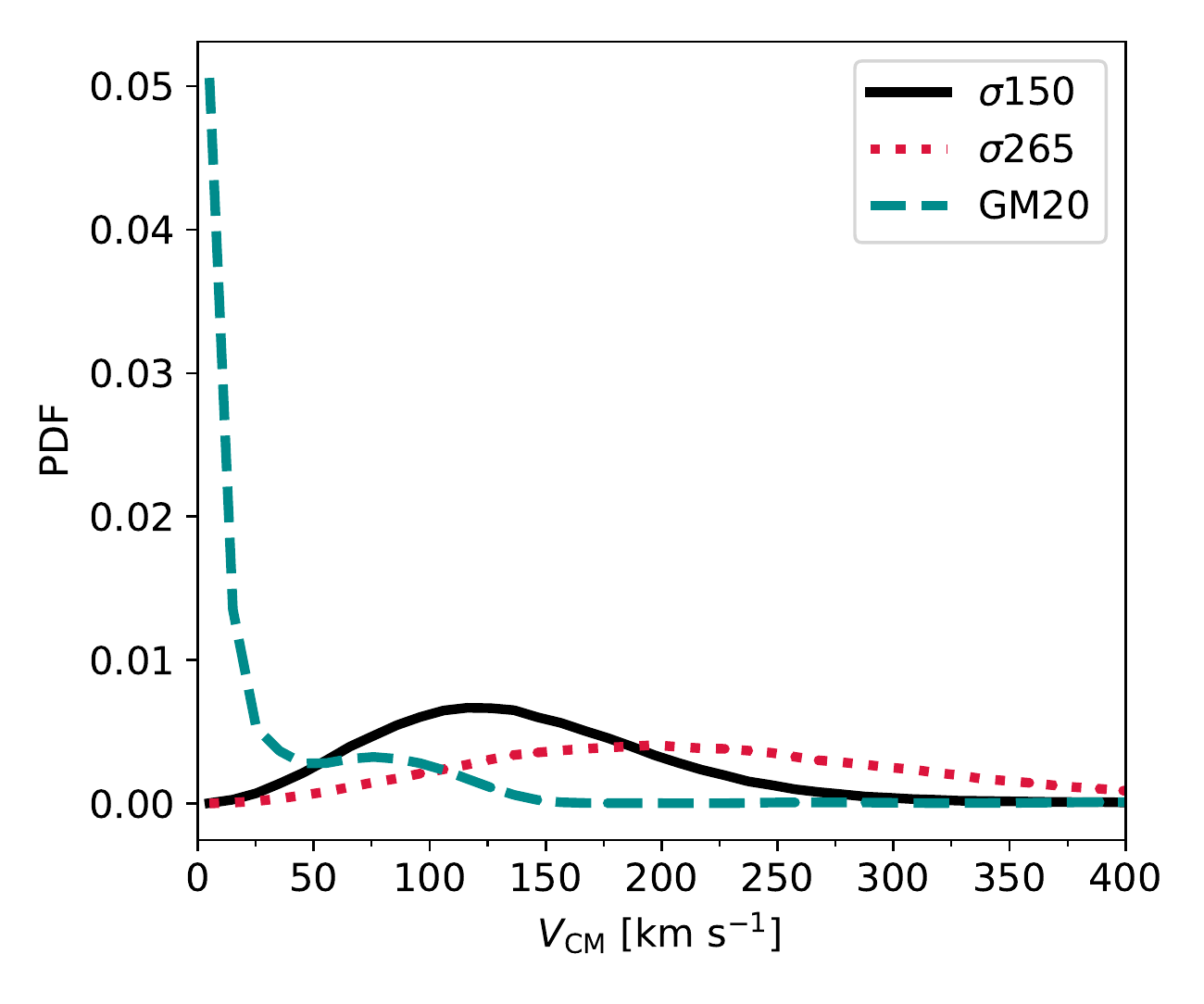}
    \caption{Probability distribution function (PDF) of the binary kick velocities in the centre of mass ($V_{\rm CM}$), for our sample of simulated BBH mergers. The centre-of-mass kick velocity takes into account both the first and the second supernova event in each binary system \protect\citep{Perna2022}. Dashed dark-cyan line: model GM20; solid black line: $\sigma{}150$; dotted red line: $\sigma{}265$. This figure only shows the kick velocity of the stellar progenitors of BBHs that merge within the lifetime of the Universe.}
    \label{fig:kick_distri}
\end{figure}

We model natal kicks of neutron stars and BHs  according to three different models, as shown in Fig.~\ref{fig:kick_distri}:
\begin{itemize}
    \item A unified kick model, in which both neutron stars and BHs receive a kick $v_{\rm kick}\propto{}m_{\rm ej}/m_{\rm rem}$, where $m_{\rm ej}$ is the mass of the ejecta and $m_{\rm rem}$ the mass of the compact remnant \citep[][hereafter GM20]{GiacobboMapelli_2020}. This model naturally produces low-kicks for electron-capture, stripped and ultra-stripped supernovae \citep{Tauris_2015,Tauris_2017}. Hereafter, we call this model GM20.
    
    \item A  model in which compact-object kicks are drawn from a Maxwellian curve with one-dimensional root-mean-square $\sigma=265$ km s$^{-1}$, consistent with observations of Galactic pulsars \citep{Hobbs_2005}. This realistically represents the upper limit for BH natal kicks. Hereafter, we name this model $\sigma{}265$.
    
    \item A model in which compact-object kicks are drawn from a Maxwellian curve with $\sigma=150$ km s$^{-1}$. This value of $\sigma$ is more similar to what suggested from indirect measurements of Galactic BH kicks \citep[e.g.,][]{Repetto_2017,Atri_2019}. Hereafter, we refer to this model as $\sigma{}150$.
\end{itemize} 

For more details about {\sc mobse}, see \cite{GiacobboMapelli_2018}. {\sc mobse} is an open-source code and can be downloaded from \href{https://gitlab.com/micmap/mobse_open}{this link}.

\subsection{Spin magnitude}
\label{sec:spins}

We have implemented four models for the spin magnitude in {\sc mobse}, the first three from \cite{Belczynski_2020}, and the fourth from \cite{Bouffanais_2019}. Given the large uncertainties on angular momentum transport, we do not claim that these four models are a complete description of the underlying physics: our models must be regarded as toy models, which bracket the current uncertainties on BH spins. 

\subsubsection{Geneva (G) model}
In the Geneva (hereafer, G) model, the dimensionless natal spin magnitude of a BH ($\chi{}$) can be approximated as:
\begin{equation}
\chi{} = \begin{cases} 
	0.85   &  M_\textup{CO} \leq m_{1} \\
	a\,{}M_\textup{CO} + b &  m_{1} < M_\textup{CO} < m_{2} \\
	a_\textup{low} & M_\textup{CO}, \geq m_{2}
\end{cases}
\label{eq:Gmodel}
\end{equation}
where $a = -0.088$ for all models, $M_{\rm CO}$ is the final carbon-oxygen mass of the progenitor star,   while the values of $b$, $m_1$, $m_2$, and $a_{\rm low}$ depend on metallicity, as indicated in Table~\ref{tab:Gspins}. This model springs from a fit by \cite{Belczynski_2020} to some evolutionary tracks by the Geneva group \citep{Ekstroem_2012}, in which angular momentum transport is relatively inefficient.

\begin{table}
\begin{tabular}{ccccc}
\hline{}
 $b$ & $m_1\,{}({\rm M}_\odot)$ & $m_2\,{}({\rm M}_\odot)$ & $a_{\rm low}$ & $Z$ \\ 
 \hline{}
  2.258  & 16.0 & 24.2 & 0.13 & $\ge{}0.010$\\
  3.578 & 31.0 & 37.8 & 0.25 & $[0.004,\,{}0.010)$\\
  2.434 & 18.0 & 27.7 & 0.0  & $[0.0012,0.004)$\\
  3.666 & 32.0 & 38.8 & 0.25 & $<0.0012$\\
 \hline{}
\end{tabular}
\caption{Parameters adopted in  model G. See Eq.~\ref{eq:Gmodel} for details.}
\label{tab:Gspins}
\end{table}

\subsubsection{MESA (M) model}

In the M model, we use the fits done by \cite{Belczynski_2020} to a set of stellar tracks run with the {\sc mesa} code. {\sc mesa} models the transport of angular momentum according to the Tayler-Spruit magnetic dynamo (\citealt{Spruit_2002}, see also \citealt{Cantiello_2014}). This yields a dimensionless natal BH spin
\begin{equation}
\chi{} = 
\begin{cases} 
	a_1\,{}M_\textup{CO} + b_1 & {\rm if}\,{} M_\textup{CO} \leq m_{1} \\
	a_2\,{}M_\textup{CO} + b_2 & {\rm if}\,{} M_\textup{CO}> m_{1},
\end{cases}
\label{eq:Mmodel}
\end{equation}
where $a_1$, $b_1$, and $m_1$ are given in Table~\ref{tab:Mspins}.

\begin{table}
\begin{tabular}{cccccc}
\hline{}
 $a_1$ & $b_1$  & $a_2$ & $b_2$ & $m_1\,{}({\rm M}_\odot)$ & $Z$ \\ 
 \hline{}
 $-0.0016$ & 0.115 & -- & -- &  $\infty$ & $\ge{}0.010$ \\
 $-0.0006$ & 0.105 & -- & -- & $\infty$ & $[0.004,\,{}0.010)$  \\
 $0.0076$  & 0.050 & $-0.0019$ & 0.165 & $12.09 $ & $[0.0012,0.004)$\\ 
 $-0.0010$ & 0.125 & -- & -- & $\infty$ &  $\le{}0.0012$\\
 \hline{}
\end{tabular}
\caption{Parameters adopted in model M. See Eq.~\ref{eq:Mmodel} for details.}
\label{tab:Mspins}
\end{table}

\subsubsection{Fuller (F) model}
\cite{Fuller_2019} predict that angular momentum transport can be even more efficient than the one predicted by the Tayler-Spruit dynamo. \cite{Belczynski_2020} summarize the results of the model by \cite{Fuller_2019} simply as
	$\chi{}= 0.01$ for all single stars and metallicities.

\subsubsection{Maxwellian model (Max)}

Finally, we also introduce a toy model in which we represent the spin of a BH as a random number drawn from a Maxwellian curve with one-dimensional root-means square $\sigma{}_\chi=0.1$ and truncated to $\chi_{\rm max}=1.0$. This model has been first introduced by \cite{Bouffanais_2019}, 
because it is a good match to the distribution arising from LVK data \citep[e.g.,][]{gwtc1,gwtc2,gwtc3_population}. Hereafter, we will indicate this Maxwellian toy model as Max, for brevity.


\subsection{Tidal spin up}
\label{sec:spinup}

The progenitor star of the second-born BH can be substantially spun-up by tidal interactions. In the scenario  explored by \cite{Bavera_2020}, a common-envelope or an efficient stable mass transfer episode can lead to the formation of a BH--WR binary system, in which the WR star is the result of mass stripping. The orbital period of this BH--WR binary system can be sufficiently short to lead to efficient tidal synchronisation and spin-orbit coupling. The WR star is then efficiently spun-up.  If the WR star then collapses to a BH directly, the final spin of the BH will retain the imprint of the final WR spin.

Based on the simulations by \cite{Bavera_2020}, \cite{Bavera_2021} derive a fitting formula to describe  the spin-up of the WR star and the final spin of the second-born BH:
\begin{equation}
\chi{} = 
\begin{cases} 
	\alpha_{\rm WR} \log_{10}^2{(P/[\textup{day}])} + \beta_{\rm WR}\,{} \log_{10}{(P/\textup{day})} & {\rm if} P \leq 1 \,{}\textrm{d}\\
	0 & \textrm{otherwise},
\end{cases}
\label{eq:bavera1}
\end{equation}
where $P$ is the orbital period of the BH--WR sytem, $\alpha_{\rm WR} = f\left(M_{\rm WR}, c^{\alpha}_{1},c^{\alpha}_{2},c^{\alpha}_{3}\right)$ and $\beta_{\rm WR}= f\left(M_{WR}, c^{\beta}_{1},c^{\beta}_{2},c^{\beta}_{3}\right)$. In this definition,  
\begin{equation}
	f\left(M_{\rm{WR}},c_1,c_2,c_3\right) = \frac{-c_1}{c_2 + \exp{\left(-c_3 M_{\rm{WR}}/[{\rm M}_\odot]\right)}},
\end{equation}
where $M_{\rm WR}$ is the mass of the WR star, while the coefficients $c_1$, $c_2$ and $c_3$ 
have been determined through non-linear least-square minimization and can be found in \cite{Bavera_2021}. 

In {\sc mobse}, we can use these fits for the spin of the second-born BH, while still adopting one of the models presented in the previous subsections (G, M, F, and  Max) for the first-born BH. 

\subsection{Spin orientation}

We assume that natal kicks are the only source of misalignment between the orbital angular momentum vector of the binary system and the direction of BH spins \citep{Rodriguez_2016,Gerosa_2018}. Furthermore, we conservatively assume that accretion onto the first-born BH cannot change the direction of its spin \citep{Maccarone_2007}. For simplicity, we also neglect the spin-flip process recently described by \citep{Stegmann_2021}. Under such assumptions, we can derive the angle between the direction of the spins of the two compact objects  and that of the orbital angular momentum of the binary system as \citep{Gerosa_2013,Rodriguez_2016}
\begin{equation}
  \cos{\delta{}}=\cos{(\nu{}_1)}\,{}\cos{(\nu{}_2)}+\sin{(\nu{}_1)}\,{}\sin{(\nu{}_2)}\,{}\cos{(\phi{})},
\end{equation}
where $\nu{}_i$  is the angle between the new (${\vec L}_{\rm new}$) and the old (${\vec L}_{\rm old}$) orbital angular momentum after a supernova ($i=1,\,{}2$ corresponding to the first and second supernova), so that $\cos{(\nu{})}={\vec L}_{\rm new}\cdot{}{\vec L}_{\rm old}/(L_{\rm new}\,{}L_{\rm old})$, while $\phi$ is the phase of the projection of the orbital angular momentum  into the orbital plane.

\subsection{Setup of {\sc mobse} runs}
\label{sec:sims}

\begin{table}
\begin{tabular}{lccl}
\hline{}
 Model Name & Spin Magnitude\footnotesize{$^a$} & B21\footnotesize{$^{b}$} & Kick Model\footnotesize{$^{c}$}\\ 
 \hline{}
 G     & Geneva (G) & no & GM20, $\sigma{}265$, $\sigma{}150$ \\
 G\_B21 & Geneva (G) & yes & GM20, $\sigma{}265$, $\sigma{}150$ \vspace{0.1cm}\\
 
 M & MESA (M) & no & GM20, $\sigma{}265$, $\sigma{}150$ \\
 M\_B21 & MESA (M) & yes & GM20, $\sigma{}265$, $\sigma{}150$\vspace{0.1cm} \\
 
 F & Fuller (F) & no & GM20, $\sigma{}265$, $\sigma{}150$ \\
  F\_B21 & Fuller (F) & yes & GM20, $\sigma{}265$, $\sigma{}150$ \vspace{0.1cm}\\

 Max & Maxwellian (Max) & no & GM20, $\sigma{}265$, $\sigma{}150$\\
 Max\_B21 & Maxwellian (Max) & yes & GM20, $\sigma{}265$, $\sigma{}150$\vspace{0.1cm}\\
 \hline{}
\end{tabular}
\caption{Description of the runs performed for this work. $^{a}$Model for the spin magnitude (Section~\ref{sec:spins}). $^{b}$Correction of the spin magnitude accounting for tidal spin up, as described in B21  (Section~\ref{sec:spinup}). $^{c}$Model for the natal kick (Section~\ref{sec:mobse}).}
\label{tab:runs}
\end{table}

Hereafter, we consider eight possible models for the spins (see also Table~\ref{tab:runs}): 
\begin{itemize}
    \item the first four models (hereafter, G, M, F, and Max) adopt the Geneva, Mesa, Fuller and Maxwellian models for both the first- and second-born BHs,
    
    \item the other four models (hereafter, G\_B21, M\_B21, F\_B21, and Max\_B21) adopt the fits by \cite{Bavera_2021} for the second-born BH and the  Geneva, Mesa, Fuller and Maxwellian  models  for the first-born BH.
\end{itemize}
For each of these eight spin models we consider three different kick models: the GM20, $\sigma{}265$, and $\sigma{}150$ models discussed in Section~\ref{sec:mobse}.

Finally, for each of these 24 models, we considered 12 metallicities ($Z=0.0002$, 0.0004, 0.0008, 0.0012, 0.0016, 0.002, 0.004, 0.006, 0.008, 0.012, 0.016, and 0.02). For each metallicity, we ran $10^7$ ($2\times{}10^7$) binary systems if $Z\leq{}0.002$ ($Z\geq{}0.004$). Hence, for each model we ran $1.8\times{}10^8$ binary systems, for a total of $4.32\times{}10^9$ binary systems encompassing the eight models.

We sampled the initial conditions for each binary system  as follows. We have randomly drawn the zero-age main sequence mass of the primary stars
from a Kroupa \citep{Kroupa_2001} initial mass function in the range $5-150$  M$_\odot$. The initial orbital
parameters (semi-major axis, orbital eccentricity and mass ratio)
of binary stars have been randomly drawn as already described in \cite{Santoliquido_2021}. In particular, we derive the mass ratios
$q \equiv{}  m_2 / m_1$ (with $m_2\leq{}m_1$) as $\mathcal{F} (q) \propto q^{ -0.1}$ with $q \in [0.1,\,{} 1]$, the orbital period $P$ from $\mathcal{F}(\Pi)\propto{}-0.55$ with $\Pi = \log_{10}{(P/{\rm d})} \in [0.15,\,{} 5.5]$ and the eccentricity $e$ from $\mathcal{F}(e)\propto{}e ^{-0.42}$ with $0\leq{}e\leq{}0.9$ \citep{Sana_2012}.

As to the main binary evolution parameters, here we use $\alpha=1$ for common envelope, while the parameter $\lambda{}$ depends on the stellar structure as described in \cite{Claeys_2014}. The other binary evolution parameters are set-up as described in \cite{Santoliquido_2021}.

\subsection{Merger rate density}
\label{sec:cosmorate}

We estimate the evolution of BBH mergers with redshift by using our semi-analytic code \cosmorate{} \citep{Santoliquido_2020,Santoliquido_2021}. With \cosmorate{}, we convolve our \mobse{} catalogues (Section~\ref{sec:sims}) with  an observation-based  metallicity-dependent  star formation rate (SFR) density evolution of the Universe, SFRD$(z,Z)$, in order to estimate the merger rate density of BBHs as 
\begin{equation}
\label{eq:mrd}
    \mathcal{R}_{\rm BBH}(z) = \int_{z_{{\rm{max}}}}^{z}\left[\int_{Z_{{\rm{min}}}}^{Z_{{\rm{max}}}} {\rm{SFRD}}(z',Z)\,{} 
    \mathcal{F}(z',z,Z) \,{}{\rm{d}}Z\right]\,{} \frac{{{\rm d}t(z')}}{{\rm{d}}z'}\,{}{\rm{d}}z',
\end{equation}
where 
\begin{equation}
\frac{{\rm{d}}t(z')}{{\rm{d}}z'} = [H_{0}\,{}(1+z')]^{-1}\,{}[(1+z')^3\Omega_{M}+ \Omega_\Lambda]^{-1/2}.
\end{equation}
In the above equation, $H_0$ is the Hubble constant, $\Omega_M$ and $\Omega_\Lambda$ are the matter and energy density, respectively. We adopt the values in \cite{Planck2018}. The term $\mathcal{F}(z',z,Z)$ is given by:
\begin{equation}
\mathcal{F}(z',z,Z) = \frac{1}{\mathcal{M}_{{\rm{TOT}}}(Z)}\frac{{\rm{d}}\mathcal{N}(z',z, Z)}{{\rm{d}}t(z)},
\end{equation}
where $\mathcal{M}_{{\rm{TOT}}}(Z)$ is the total simulated initial stellar mass, and  ${{\rm{d}}\mathcal{N}(z',z, Z)/{\rm{d}}}t(z)$ is the rate of BBHs forming from stars with initial metallicity $Z$ at redshift $z'$ and merging at $z$, extracted from our \mobse{} catalogues. 
In \cosmorate{},    ${\rm{SFRD}}(z,Z)$  is given by 
\begin{equation}
{\rm{SFRD}}(z',Z) = \psi(z')\,{}p(z',Z),
\end{equation}
where $\psi(z')$ is the cosmic SFR density at formation redshift $z'$, and $p(z',Z)$ is the log-normal distribution of metallicities $Z$ at fixed formation redshift $z'$, with average $\mu(z')$ and spread $\sigma_{Z}$. Here, we take both $\psi{}(z)$ and $\mu{}(z)$ from \cite{Madau_2017}. 
 Finally, we assume a metallicity spread $\sigma_Z = 0.3$.

\subsection{Hyper-parametric model description}
\label{sec:hyper}

For each of our models (Table~\ref{tab:runs}), described by their hyper-parameters $\lambda$,  we predict the distributions of  BBH mergers
\begin{equation}
    \frac{\mathrm{d}N}{\mathrm{d}\theta}(\lambda) = N_\lambda \,{}p(\theta|\lambda),
\end{equation}
where $\theta$ are the merger parameters, and $N_\lambda$ is the total number of mergers predicted by the model. Assuming an instrumental horizon redshift $z_{\rm max} =1.5$, $N_\lambda$ can be calculated as
\begin{equation}
    N_\lambda = \int_0^{z_{\rm max}} \mathcal{R}(z)\,{}\frac{{\rm d}V_{\rm c}}{{\rm d}z}\,{} \frac{T_{\rm obs}}{(1+z)}\,{}{\rm d}z,
    \label{eq:Nlambda}
\end{equation}
where $\frac{{\rm d}V_{\rm c}}{{\rm d}z}$ is the comoving volume and $T_{\rm obs}$ the observation duration. 

To model the population of merging BBHs, we have chosen five observable parameters $\theta = \{\mathcal{M}_{\rm c},\,{} q,\,{} z, \,{}\chi_{\rm eff},\,{}\chi_{\rm p} \}$, where  $\mathcal{M}_{\rm c} = (m_1\,{}m_2)^{3/5}/(m_1+m_2)^{1/5}$ is the chirp mass in the source frame with $m_1$ ($m_2$) the masses of the primary (secondary) BH of the binary, $q= m_2/m_1$. 
and $z$ is the redshift of the merger. In addition, we used two spin parameters: the effective spin ($\chi_{\rm eff}$) and the precessing spin ($\chi_{\rm p}$). The effective spin $\chi_{\rm eff}$ is the mass-weighted projection of the two individual BH spins on the binary orbital angular momentum $\vec L$
\begin{equation}
\chi_{\rm eff} = \frac{(\vec \chi_1 + q\,{}\vec \chi_2)}{1+q} \cdot{} \frac{\vec L}{L}, 
\end{equation}
where  $\vec \chi_{1,2} = {\vec s_{1,2}\,{}c/(G\,{}m_{1,2}^2)}$ is the dimensionless spin  parameter of the two BHs. 
 The precessing spin $\chi_{\rm p}$  
is defined as
\begin{equation}
\chi_{\rm p} = {\rm max}\left(\chi_{1,\perp},\,{} A\, \chi_{2,\perp}\right), 
\end{equation}
where 
$\chi_{1,\perp}$ ($\chi_{2,\perp}$) is the spin component of the primary (secondary) BH perpendicular to the orbital angular momentum vector $\vec{L}$, and $A = \left({4\,{}q+3}\right)\,{}q/\left({4+3\,{}q}\right)$.

To compute the distributions $p(\theta|\lambda)$, we constructed a catalogue of $10^6$ sources for all possible combinations of hyper-parameters $\lambda$, using the merger rate density and the metallicity given by \cosmorate{}. From these catalogues we derived continuous estimations of $p(\theta|\lambda)$ by making use of a Gaussian kernel density estimation assuming a bandwidth of 0.15.





\begin{figure}
    \includegraphics[width=\columnwidth]{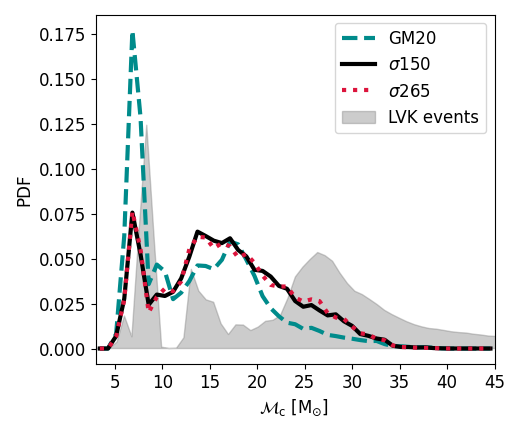}
    \caption{Predicted detectable distribution of chirp mass, for each kick model: GM20 (solid dark-cyan line), $\sigma{}150$ (dotted black line) and $\sigma265$ (dashed red line). For detectable distribution we mean the distribution of simulated BBHs with sufficiently high signal-to-noise ratio (Section~\ref{sec:analysis}). The shaded gray area is the distribution we obtain by stacking the posterior samples of the 59 confident detections from GWTC-3 (Appendix~\ref{sec:events}).}
    \label{fig:Mc_distri}
\end{figure}


\begin{figure*}
    \includegraphics[width=2\columnwidth]{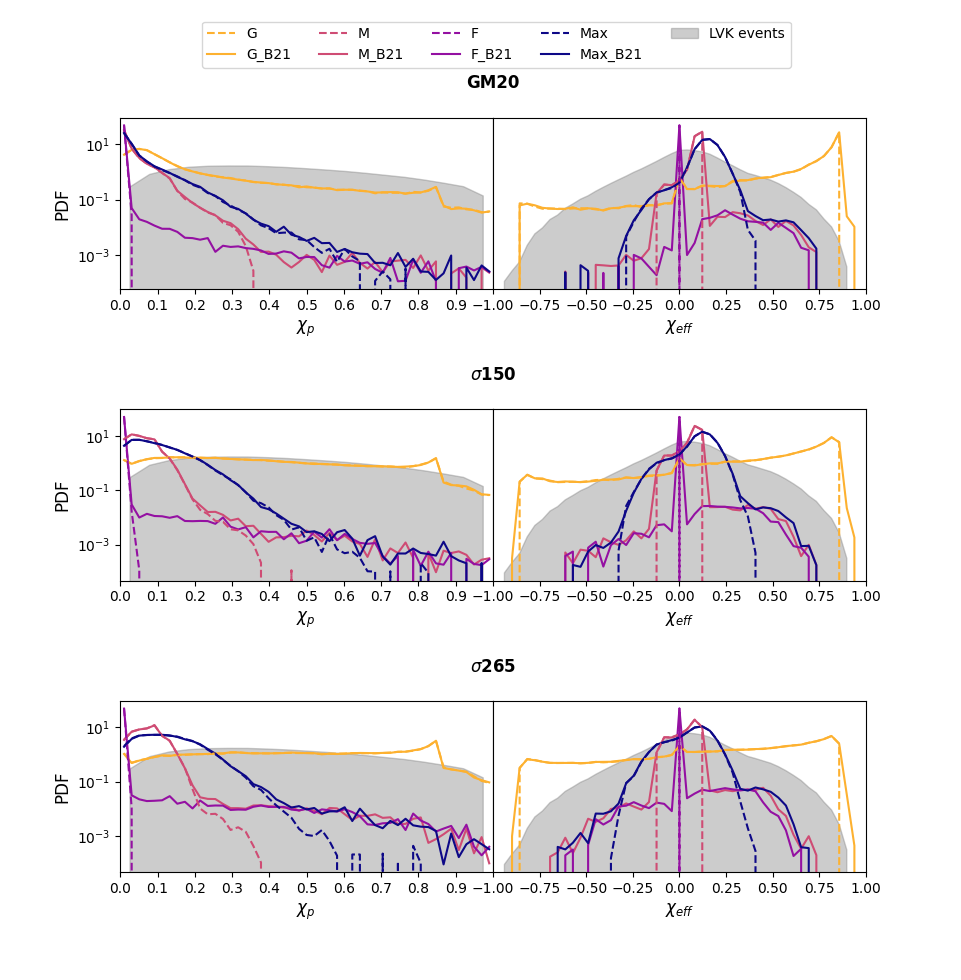}
    \caption{Predicted detectable distribution of $\chi_{\rm p}$ (left) and $\chi_{\rm eff}$ (right) for all of our models. Different colours refer to the spin model: G, M, F and Max. Solid (dashed) lines include (do not include) the tidal spin-up model by B21. From top to bottom: GM20, $\sigma{}150$, and $\sigma{}265$. The shaded gray areas are the distributions we obtain by stacking the posterior samples of the 59 confident detections from GWTC-3 (Appendix~\ref{sec:events}).}
    \label{fig:chi_distri}
\end{figure*}

\section{Hierarchical Bayesian inference}\label{sec:analysis}

 Given a set $\mathcal{H}=\{h^k\}_{k=1}^{N_{\rm obs}} $ of $N_{\rm obs}$ GW observations, the posterior distribution of a set of hyper-parameters $\lambda$ associated to an astrophysical model can be described as an in-homogeneous Poisson distribution \citep[e.g.,][]{Loredo_2004,Mandel_2019,Thrane_2019,Bouffanais_2019,Bouffanais_2021a,Bouffanais_2021b}:
\begin{equation}
    p(\lambda, N_\lambda|\mathcal{H}) = e^{-\mu_\lambda}\,\pi(\lambda, N_\lambda)\prod^{N_{\rm obs}}_{k=1}N_\lambda\int\mathcal{L}^k(h^k|\theta)\,p(\theta|\lambda)\,\mathrm{d}\theta,
    \label{eq:p1}
\end{equation}
where $N_{\rm obs}$ is the number of events observed by the LVK,  with an ensemble of parameters $\theta$, $N_\lambda$ is the number of predicted mergers by the model (as calculated in eq.~\ref{eq:Nlambda}), $\mu_\lambda$ the number of predicted observations given a model and a detector, $\pi(\lambda,\,{} N_\lambda)$ are the prior distributions on $\lambda$ and $N_\lambda$, and $\mathcal{L}^k(\{h\}^k|\theta)$ is the likelihood of the $k^{\rm th}$ observation.

 The predicted number of events $\mu_\lambda$ can be written in terms of detection efficiency $\beta(\lambda)$ for a given model:
\begin{equation}
    \mu_\lambda = N_\lambda\,\beta(\lambda) , \quad \mathrm{with} \quad \beta(\lambda) = \int_{\theta}p(\theta|\lambda)\,{}p_{\rm det}(\theta)\,{}\mathrm{d}\theta,
    \label{eq:p2}
\end{equation}
where $p_{\rm det}(\theta{})$ is the detection probability for a set of parameters $\theta$. 
This probability can be inferred by computing the optimal signal to noise ratio (SNR) of the sources and comparing it to a detection threshold. In our case we chose as reference a threshold $\rho_{\rm thr} = 8$ in the LIGO Livingston detector, for which we approximated the sensitivity using the measurements for the three runs separately \citep{LIGO_2010, LIGO_2016, Wysocki_2018}.   
The values for the event's log-likelihood were derived from the posterior and prior samples released by the LVK. 
Hence, the integral in eq.~\ref{eq:p1} is approximated with a Monte Carlo approach as
\begin{equation}
    \mathcal{I}^k = \int \mathcal{L}^k(h^k|\theta)\,{}p(\theta|\lambda)\,{}\mathrm{d}\theta \approx \frac{1}{N^k_s}\sum^{N^k_s}_{i=1}\frac{p(\theta_i^k|\lambda)}{\pi^k(\theta^k_i)},
    \label{eq:indiv_logLike}
\end{equation}
where $\theta^k_i$ is the $i^{\rm th}$ posterior sample of the $k^{\rm th}$ detection and $N_s^k$ is the total number of posterior samples for the $k^{\rm th}$ detection. To compute the prior term in the denominator, we also used Gaussian kernel density estimation.

Finally, we can also choose to neglect the information coming from the number of sources predicted by the model when estimating the posterior distribution. By doing so, we can have some insights on the impact of the rate on the analysis. In practice, this can be done by marginalising eq.~\ref{eq:p1} over $N_\lambda$ using a prior $\pi(N_\lambda)\sim 1/N_\lambda$ \citep{Fishbach_2018}, which yields to the following expression for a model log-likelihood

\begin{equation}
    \mathcal{L} = p(\lambda|\{h^k\})\sim \pi(\lambda)\,{}\prod_{k=1}^{N_{\rm obs}}\left[\frac{ \mathcal{I}^k}{\beta(\lambda)}\right].
    \label{eq:p4}
\end{equation}

We adopted the formalism described in eqs.~\ref{eq:p1}--\ref{eq:p4} to perform a hierarchical Bayesian inference to compare the astrophysical models presented Sec.~\ref{sec:models} with the third gravitational-wave transient catalogue (GWTC-3), the most updated catalogue of gravitational-wave events from the LVK \citep{gwtc3_catalogue, gwtc3_population}. GWTC-3 contains 90 event candidates with probability of astrophysical origin $p_{\rm astro}>0.5$. From GWTC-3, we extract 59 confident detections of BBHs with a false alarm rate ${\rm FAR}<0.25$~yr$^{-1}$. In this sub-sample, we do not include binary neutron stars and neutron star -- BH systems, and we also exclude the other BBH candidates with an higher FAR. Our chosen FAR threshold  ensures a sufficiently pure sample for our analysis \citep{gwtc3_population}. A list of the events used in this study is available in Appendix \ref{sec:events}. For the observable parameters $\theta$, we use the choice described in Section~\ref{sec:hyper}, namely $\theta = \{\mathcal{M}_{\rm c},\,{} q,\,{} z, \,{}\chi_{\rm eff},\,{}\chi_{\rm p} \}$. 

\section{Results}\label{sec:results}


\subsection{Chirp mass}\label{sec:chirp}

The chirp mass distribution (Fig.~\ref{fig:Mc_distri}) does not depend on the spin model, by construction. Therefore, we only show different natal kicks. Models $\sigma{}150$ and $\sigma{}265$ show a similar distribution of chirp masses with two peaks of similar importance, one at $\mathcal{M}_{\rm c}\approx{8}$~M$_\odot$ and the other (broader) peak  at $\mathcal{M}_{\rm c}\approx{15}$~M$_\odot$. In contrast, model GM20 has a much stronger preference for low-mass BHs, with a dominant peak at $\mathcal{M}_{\rm c}\approx{8}$~M$_\odot$. The reason for this difference is that  all BHs in tight binary systems receive slow natal kicks in model
GM20 (Fig.~\ref{fig:kick_distri}). This happens because stars in tight binary systems lose their envelope during mass transfer episodes; hence, the mass of supernova ejecta  ($m_{\rm ej}$) is small, triggering low kicks in model GM20.  

Figure~\ref{fig:Mc_distri} also compares the detectable distribution of our models  with the stacked posterior samples from the confident BBH detections in GWTC-3. This figure highlights two main differences  between the population synthesis models and the posterior samples: the peak at $\mathcal{M}_{\rm c}\approx{15}$ M$_\odot$ is stronger in the models than it is in the data, while the data present a more significant excess at $\mathcal{M}_{\rm c}\approx{25-30}$ M$_\odot$ than the models. Finally, the peak at $\mathcal{M}_{\rm c}\approx{9}$ M$_\odot$ in the data approximately matches the peak at $\mathcal{M}_{\rm c}\approx{8}$ M$_\odot$ in the models. The main features of our population synthesis models (in particular, the peaks at $\mathcal{M}_{\rm c}\approx{8-10}$ M$_\odot$ and $\mathcal{M}_{\rm c}\approx{15-20}$ M$_\odot$) are also common to other population-synthesis models \citep[e.g.,][]{Belczynski_2020,Vanson_2022} and mostly spring  from the core-collapse SN prescriptions by \cite{Fryer_2012}. Alternative core-collapse SN models \citep[e.g.,][]{Mapelli_2020,Mandel_2021,Patton_2022,Olejak_2022} produce different features and deserve further investigation (Iorio et al., in prep.).

\subsection{Spin parameters}\label{sec:spin}

Figure~\ref{fig:chi_distri} shows the detectable distribution of spin parameters $\chi_{\rm p}$ and $\chi_{\rm eff}$ for all of our models. By construction, large spins are much more common in models G and G\_B21, while models F and F\_B21 have a strong predominance of vanishingly small spins. Models M, M\_B21, Max and Max\_B21 are intermediate between the other two extreme models. 

Including or not the correction by B21 has negligible impact on the distribution of $\chi_{\rm p}$ and $\chi_{\rm eff}$ for models G, because of the predominance of large spin magnitudes. In contrast, introducing the spin-up correction by B21 has a key impact on models F, because it is the only way to account for mild to large spins in these models. The correction by B21 is important also for models M and Max, being responsible for the large-spin wings.

Finally, our model with slow kicks (GM20) results in a distribution of $\chi_{\rm p}$ that is more peaked at zero (for models G, M and Max) with respect to the other two kick models ($\sigma{}150$ and $\sigma{}265$). In fact, the supernova kicks in model GM20 are not large enough to appreciably misalign BH spins (see Fig.~\ref{fig:kick_distri}). 

A similar effect is visible in the distribution of $\chi_{\rm eff}$: model $\sigma{}265$ produces a distribution of $\chi_{\rm eff}$ that is less asymmetric about the zero with respect to models $\sigma{}150$ and especially GM20.

\subsection{Model Selection} 
\label{sec:model_sel}

Figure~\ref{fig:results} and Table~\ref{tab:results} report the values of the log-likelihood $\log\mathcal{L}$ defined in Eq.~\ref{eq:p4}.
We can quantify the difference between two models A and B by computing the average absolute difference in percentage 
\begin{equation}
    \Delta{\rm log}\mathcal{L}({\rm A,\,B}) = \left<\frac{2\left|{\rm log}\mathcal{L}_i^{\rm A}-{\rm log}\mathcal{L}_i^{\rm B}\right|}{{\rm log}\mathcal{L}_i^{\rm A}+{\rm log}\mathcal{L}_i^{\rm B}}\right>_{var},
    \label{eq:average_diff}
\end{equation}
on the non-A,B variation $var$ ($var$ would be kick(spin) if A and B are spin(kick) models). For example to compare the two models G and G\_B21, A and B become G\_B21 and G and $var = \{{\rm GM}20,\, \sigma150, \,\sigma265\}$.

The tidal spin-up mechanism (B21) affects the spin of 
a small part of the population of each model (Fig.~\ref{fig:chi_distri}). However, it improves  the likelihood of the F and M models significantly (e.g., $\Delta{\rm log}\mathcal{L}({\rm M\_B21},\,{}{\rm M}) =89\%$, Table~\ref{tab:results}). 
This improvement of the log-likelihood can be explained by the presence of higher values of  $\chi_{\rm p}$ and $\chi_{\rm eff}$ in the distribution of populations M\_B21 and F\_B21 compared to M and F (Fig.~\ref{fig:chi_distri}). 

The F model yields $\mathcal{L}({\rm F})=-\infty{}$ if we do not include the tidal spin-up correction, regardless of the kick model.  
This indicates that the LVK data do not support vanishingly small BH spins for the
entire BBH population. 
However, it is sufficient to inject a tiny sub-population of spinning
BHs, by switching on the B21 correction, and the F model becomes
one of the best considered models. In fact, the F\_B21 models only
includes 0.4\% of BHs with $\chi > 0.01$ and achieves $\log{\mathcal{L}} > 200$ (for
spin models $\sigma{150}$ and $\sigma{265}$).

The G and G\_B21 spin models  exhibit lower log-likelihood values than the others for all kicks models: ${\rm log}\mathcal{L}	\leqslant 150$ for $\sigma{150}$ and $\sigma265$, and ${\rm log}\mathcal{L}	< 0$ for GM20. This happens because 
the distribution of $\chi_{\rm eff}$  has non-negligible support for extreme values $\chi_{\rm eff}<-0.5$ and $\chi_{\rm eff}>0.5$   (Fig.~\ref{fig:chi_distri}). 

The kick models 
$\sigma{150}$ and $\sigma265$ show similar results ($\Delta{\rm log}\mathcal{L}({\rm \sigma{150},{}\sigma265})<3\%)$ for every spin assumptions. Also, for all spin assumptions, the GM20 kick model scores a significantly lower likelihood than the other models $\sigma{150}$ and $\sigma265$ with $\Delta{\rm log}\mathcal{L}({\rm \sigma150,{}{\rm GM20}})\,\sim\, \Delta{\rm log}\mathcal{L}({\rm \sigma265,{}GM20})\,\sim\,$150\%.
This result can be explained by the high peak of model GM20 at low chirp masses ($\mathcal{M}_{\rm c}\sim 8 {\rm M}_\odot$, see Sec.\ref{sec:chirp} and Fig.\ref{fig:Mc_distri}) and  by the low value of $\chi{}_{\rm p}$  compared to the other kick models (Fig.~\ref{fig:chi_distri}).




Models Max and Max\_B21 are  possibly the best match to the data, but this is not surprising, because they were built as a toy model to visually match the data. 
Among the astrophysically-motivated models (i.e., after excluding the Max model), M, M\_B21 and F\_B21 (with kick models $\sigma{150}$ and $\sigma{}265$) are the most favoured by the data. 
This might be interpreted as a support for the Tayler-Spruit instability mechanism (adopted in models M) and for the tidal spin-up model by B21.

\subsection{Importance of $\chi{\rm _p}$}

\begin{figure*}
    \includegraphics[width=13cm]{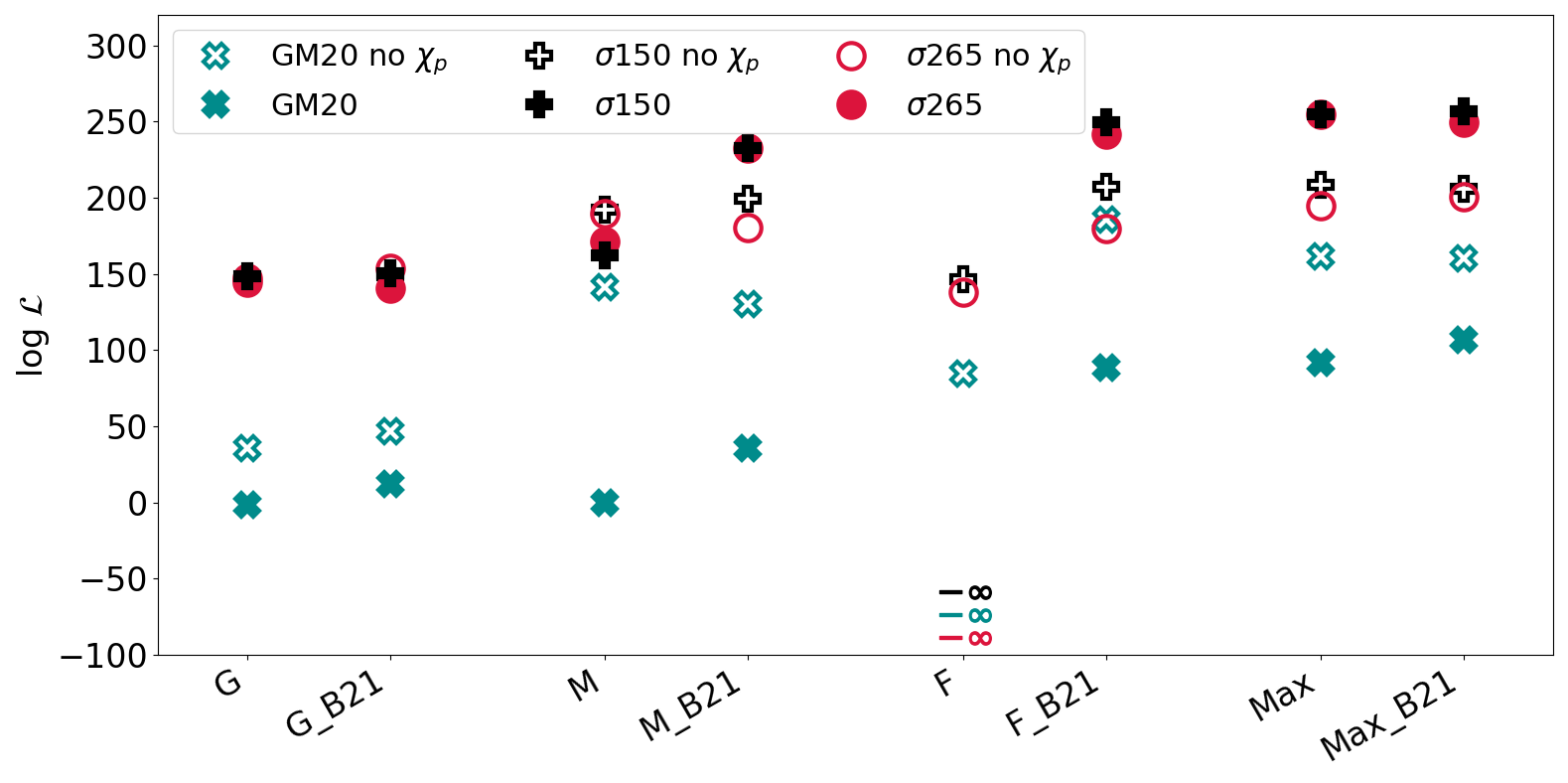}
    \caption{Values of the log-likelihood $\mathcal{L}$ defined in Eq.~\ref{eq:p4} for the four different models Geneva (G), MESA (M), Fuller (F), and Maxwellian (Max), with/without the tidal spin-up mechanism (B21). Blue crosses: GM20; dark pluses: $\sigma{}150$; red circles: $\sigma{}265$.}
    \label{fig:results}
\end{figure*}

The $\chi{\rm _p}$ parameter 
encodes information on the spin component in the orbital plane. Its impact on gravitational-wave signals is much lower than that of $\chi_{\rm eff}$, and therefore its measurement is less precise. 
To understand the impact of $\chi_{\rm p}$ on our results, we re-ran the analysis without this parameter. The results are shown in Table~\ref{tab:results_nochip} and  in Fig.~\ref{fig:results} with empty markers. 
Fig.~\ref{fig:results} shows that, if we do not  include $\chi_{\rm p}$, the models M and M\_B21 have almost the same log-likelihood, and even the F model yields a  positive log-likelihood.
Furthermore, the analysis without $\chi_{\rm p}$ results in significantly larger values of $\mathcal{L}$ for the kick model GM20.
Our results demonstrate that the measured $\chi_{\rm p}$ of GWTC-3 BBHs  carries substantial information, despite the large uncertainties.

\begin{table}
\centering
\caption{Log-likelihood $\mathcal{L}$ (Eq.~\ref{eq:p4}) estimated with five merger parameters $\theta = \left\{\mathcal{M}_{\rm c}\,, z\,, \chi_{\rm eff}\,, q\,,\chi_{\rm p}  \right\}$.}
\label{tab:results}
\begin{tabular}{lccc}
\hline
Model Name & GM20 & $\sigma$150 & $\sigma$265 \\ \hline
G                & -1    & 149                        & 145                \\
G\_B21           & -12    & 150                         & 141         \vspace{0.1cm}       \\
M                  & 0   & 162                        & 171                \\
M\_B21            & 36    & 232                        & 232         \vspace{0.1cm}       \\
F                   & -$\infty$     & -$\infty$            & -$\infty$            \\
F\_B21             & 88   & 250                       & 242         \vspace{0.1cm}       \\
Max               & 92   & 255                        & 254                \\
Max\_B21            & 106  & 257                       & 250                \\ \hline
\end{tabular}
\end{table}

\begin{table}
\centering
\caption{Log-likelihood $\mathcal{L}$ (Eq.~\ref{eq:p4}) estimated with four merger parameters $\theta = \left\{\mathcal{M}_{\rm c}\,, z\,, \chi_{\rm eff}\,, q \right\}$. Here, we ignore $\chi_{\rm p}$.}
\label{tab:results_nochip}
\begin{tabular}{lccc}
\hline
Model Name & GM20 & $\sigma$150 & $\sigma$265 \\ \hline
G                  & 35   & 146                       & 147                \\
G\_B21             & 47  & 149                         & 154         \vspace{0.1cm}       \\
M                 & 141   & 192                         & 190                \\
M\_B21             & 130   & 199                        & 180         \vspace{0.1cm}       \\
F                 & 85   & 146                & 138            \\
F\_B21             & 185   & 207                       & 180         \vspace{0.1cm}       \\
Max                & 161   & 208                       & 155                \\
Max\_B21          & 160   & 206                        & 200                \\ \hline
\end{tabular}
\end{table}

\section{Discussion}
\label{sec:conclusion}

The spin magnitude of BHs is largely uncertain, mostly because we do not fully understand angular momentum transport in massive stars. Here, we have taken 
a number of spin models bracketing the main uncertainties, we have implemented them into our population-synthesis code {\sc mobse}, and compared them against GWTC-3 data within a hierarchical Bayesian framework.

The data do not support models in which the entire BH population has vanishingly small spins (model F). This result is mainly driven by the $\chi_{\rm p}$ parameter. This is in agreement with, e.g., the complementary analysis presented in \cite{Callister_2022}. They employed a variety of complementary methods to measure the distribution of spin magnitudes and orientations of BBH mergers, and concluded that the
existence of a sub-population of BHs with vanishing spins is not required by current data. \cite{Callister_2022} find that the fraction
of non-spinning BHs can comprise up to $\sim{60 - 70}$\%
of the total population. 
In our F\_B21 models, we have $\sim{99.6}$\% of BHs with $\chi<0.01$.

Recently, \cite{Biscoveanu_2020, Roulet_2021, Galaudage_2021} and \cite{Tong_2022} claimed the existence of a sub-population of zero-spin BHs. From our analysis, we cannot exclude the existence of such sub-population, as the F model with B21 correction (F\_B21) still represents a good match of the data. Similarly to \cite{Belczynski_2020} and \cite{Gerosa_2018}, we find that models with large spins (G, G\_B21) are less favoured by the data, but they are still acceptable if we allow for large kicks.

Overall, we find a preference for large natal kicks. 
This result goes into the same direction as the work by \cite{Callister_2021}. Actually,  this preference for large natal kicks is degenerate with the adopted formation channel. Had we included the dynamical formation channel in dense star clusters, we would have added a sub-population of isotropically oriented spins (see, e.g., Figure~8 of \citealt{Mapelli_2022}). In a forthcoming study, we will extend our analysis to a multi-channel analysis. While it is unlikely that BBH mergers only originate from one single channel, adding more formation channels to a hierarchical Bayesian analysis dramatically increases the number of parameters, making it more difficult to reject some portions of the parameter space.


\section{Summary}
\label{sec:summary}


The origin of BH spins is still controversial, and angular momentum transport inside massive stars is one of the main sources of uncertainty. 
Here, we apply hierarchical Bayesian inference to derive constraints on spin models from the 59 most confident BBH merger events in GWTC-3. We consider  five parameters:  chirp mass, mass ratio, redshift, effective spin, and precessing spin. 

For model selection, we use a  set of binary population synthesis simulations spanning  different assumptions for black hole spins and natal kicks. In particular, our spin models  account for relatively inefficient (G), efficient  (Max and M), and very efficient angular-momentum transport (F). A higher efficiency of angular momentum transport is associated with lower  BH spins. In particular, model F predicts vanishingly small spins for the entire BH population. For each of our models, we also include the possibility that some BHs are tidally spun-up (B21). We considered three different natal kick models: according to models $\sigma{}265$ and $\sigma{}150$, we randomly draw the kicks from a Maxwellian curve with $\sigma{}=265$ and 150 km~s$^{-1}$, respectively; in the third model (G20), we also derive the kicks from a Maxwellian curve with $\sigma{}=265$ km~s$^{-1}$, but the kick magnitude is then modulated by the ratio between the mass of the ejecta and the mass of the BH.

We summarize our main results as follows.
\begin{itemize}
\item The data from GWTC-3 do not support models in which the entire BH population has vanishingly small spins (model F). 
\item In contrast, models in which most spins are vanishingly small, but that also include a sub-population of
tidally spun-up BHs (model F\_B21) are a good match to the data. 
\item The models in which angular momentum transport is relatively inefficient (G and G\_21) yield log-likelihood values that are much lower than models with efficient angular momentum transport (M, M\_B21, Max, and Max\_B21).
\item Models with large BH kicks  ($\sigma{}150$ and $\sigma{}265$ ) are favoured by our analysis with respect to low-kick models (G20). 
    \item Our results show that the precessing spin parameter $\chi_{\rm p}$ plays a crucial impact to constrain the spin distribution of BBH mergers.
\end{itemize}

%
%
%
%
%
%


\section*{Acknowledgements}

 MM, CP, FS and YB acknowledge financial support from the European Research  Council for the ERC Consolidator grant DEMOBLACK, under contract no.  770017. This research has made use of data or software obtained from the Gravitational Wave Open Science Center (gwosc.org), a service of LIGO Laboratory, the LIGO Scientific Collaboration, the Virgo Collaboration, and KAGRA. LIGO Laboratory and Advanced LIGO are funded by the United States National Science Foundation (NSF) as well as the Science and Technology Facilities Council (STFC) of the United Kingdom, the Max-Planck-Society (MPS), and the State of Niedersachsen/Germany for support of the construction of Advanced LIGO and construction and operation of the GEO600 detector. Additional support for Advanced LIGO was provided by the Australian Research Council. Virgo is funded, through the European Gravitational Observatory (EGO), by the French Centre National de Recherche Scientifique (CNRS), the Italian Istituto Nazionale di Fisica Nucleare (INFN) and the Dutch Nikhef, with contributions by institutions from Belgium, Germany, Greece, Hungary, Ireland, Japan, Monaco, Poland, Portugal, Spain. KAGRA is supported by Ministry of Education, Culture, Sports, Science and Technology (MEXT), Japan Society for the Promotion of Science (JSPS) in Japan; National Research Foundation (NRF) and Ministry of Science and ICT (MSIT) in Korea; Academia Sinica (AS) and National Science and Technology Council (NSTC) in Taiwan \cite{1912.11716, 2302.03676}. This research made use of \textsc{NumPy} \citep{Harris_2020}, and \textsc{SciPy} 
 \citep{SciPy_2020}. For the plots we used \textsc{Matplotlib} \citep{Hunter_2007}.

\section*{Data Availability}
The data underlying this article will be shared on reasonable request to the corresponding author. The latest public version of {\sc mobse} can be downloaded from \href{https://gitlab.com/micmap/mobse_open}{this repository}. \cosmorate{} can be downlowaded from \href{https://gitlab.com/Filippo.santoliquido/cosmo_rate_public}{this link}.



\bibliographystyle{mnras}
\bibliography{biblio} 




\appendix

\section{Sample of gravitational-wave events}
\label{sec:events}

Table \ref{tab:goldcat} lists all the gravitational-wave event candidates we used in our study. From GWTC-3, we selected all the event candidates with $p_{\rm astro}>0.9$ and FAR$< 0.25$ yr$^{-1}$, excluding the following three systems:
\begin{itemize}
    \item the binary neutron star GW170807;
    
    \item the (possible)  neutron star--BH binary system GW190814;
    
    \item the BBH GW190521 ($m_1 = 98.4^{+33.6}_{-21.7} \mathrm{M}_\odot$, $m_2 = 57.2^{+27.1}_{-30.1} \mathrm{M}_\odot$ \citealt{gwtc3_catalogue}), which can form only via dynamical interactions in our models \citep[e.g.,][]{dicarlo2019,dallamico2021,Mapelli_2021}.
\end{itemize}

\begin{table*}
\renewcommand{\arraystretch}{1.2}
\centering
\caption{Catalogue of BBH events adopted in this study. the ucertainties shown stand for the 90\% credible intervals. }
\label{tab:goldcat}
\scalebox{0.9}{
\begin{tabular}{lccccc}
\hline 
\myrowcolour
\bf{Name}  &  $\mathbf{\mathcal{M}_{\rm c}} \mathbf{[M}\mathbf{_\odot}\mathbf{]}$ &  $\mathbf{q}$ & $\mathbf{\chi_{\rm eff}}$  & $\mathbf{\chi_{\rm p}}$ & $\mathbf{z}$ \\ \hline

GW150914 & 28.6$_{-1.5}^{+1.7}$ & 0.86$_{-0.2}^{+0.12}$ & -0.01$_{-0.13}^{+0.12}$ & 0.34$_{-0.25}^{+0.45}$ & 0.09$_{-0.03}^{+0.03}$ \\ \myrowcolour
GW151012 & 15.2$_{-1.2}^{+2.1}$ & 0.59$_{-0.35}^{+0.36}$ & 0.05$_{-0.2}^{+0.31}$ & 0.33$_{-0.25}^{+0.45}$ & 0.21$_{-0.09}^{+0.09}$\\
GW151226 & 8.9$_{-0.3}^{+0.3}$ & 0.56$_{-0.33}^{+0.38}$ & 0.18$_{-0.12}^{+0.2}$ & 0.49$_{-0.32}^{+0.39}$ & 0.09$_{-0.04}^{+0.04}$\\ \myrowcolour
GW170104 & 21.4$_{-1.8}^{+2.2}$ & 0.65$_{-0.23}^{+0.3}$ & -0.04$_{-0.21}^{+0.17}$ & 0.36$_{-0.27}^{+0.42}$ & 0.2$_{-0.08}^{+0.08}$\\
GW170608 & 7.9$_{-0.2}^{+0.2}$ & 0.69$_{-0.36}^{+0.28}$ & 0.03$_{-0.07}^{+0.19}$ & 0.36$_{-0.27}^{+0.45}$ & 0.07$_{-0.02}^{+0.02}$\\ \myrowcolour
GW170729 & 35.4$_{-4.8}^{+6.5}$ & 0.68$_{-0.28}^{+0.28}$ & 0.37$_{-0.25}^{+0.21}$ & 0.44$_{-0.28}^{+0.35}$ & 0.49$_{-0.21}^{+0.19}$\\
GW170809 & 24.9$_{-1.7}^{+2.1}$ & 0.68$_{-0.24}^{+0.28}$ & 0.08$_{-0.17}^{+0.17}$ & 0.35$_{-0.26}^{+0.43}$ & 0.2$_{-0.07}^{+0.05}$\\ \myrowcolour
GW170814 & 24.1$_{-1.1}^{+1.4}$ & 0.83$_{-0.23}^{+0.15}$ & 0.07$_{-0.12}^{+0.12}$ & 0.48$_{-0.36}^{+0.41}$ & 0.12$_{-0.04}^{+0.03}$\\
GW170818 & 26.5$_{-1.7}^{+2.1}$ & 0.76$_{-0.25}^{+0.21}$ & -0.09$_{-0.21}^{+0.18}$ & 0.49$_{-0.34}^{+0.37}$ & 0.21$_{-0.07}^{+0.07}$\\ \myrowcolour
GW170823 & 29.2$_{-3.6}^{+4.6}$ & 0.74$_{-0.3}^{+0.23}$ & 0.09$_{-0.26}^{+0.22}$ & 0.42$_{-0.31}^{+0.41}$ & 0.35$_{-0.15}^{+0.15}$\\
GW190408\_181802 & 18.3$_{-1.2}^{+1.9}$ & 0.75$_{-0.24}^{+0.21}$ & -0.03$_{-0.19}^{+0.14}$ & 0.39$_{-0.31}^{+0.37}$ & 0.29$_{-0.1}^{+0.06}$\\ \myrowcolour
GW190412 & 13.3$_{-0.3}^{+0.4}$ & 0.28$_{-0.06}^{+0.12}$ & 0.25$_{-0.11}^{+0.08}$ & 0.3$_{-0.16}^{+0.19}$ & 0.15$_{-0.03}^{+0.03}$\\
GW190413\_052954 & 24.6$_{-4.1}^{+5.5}$ & 0.69$_{-0.29}^{+0.27}$ & -0.01$_{-0.34}^{+0.29}$ & 0.41$_{-0.31}^{+0.43}$ & 0.59$_{-0.24}^{+0.29}$\\ \myrowcolour
GW190413\_134308 & 33.0$_{-5.4}^{+8.2}$ & 0.69$_{-0.31}^{+0.28}$ & -0.03$_{-0.29}^{+0.25}$ & 0.56$_{-0.41}^{+0.37}$ & 0.71$_{-0.3}^{+0.31}$\\
GW190421\_213856 & 31.2$_{-4.2}^{+5.9}$ & 0.79$_{-0.3}^{+0.19}$ & -0.06$_{-0.27}^{+0.22}$ & 0.48$_{-0.36}^{+0.4}$ & 0.49$_{-0.21}^{+0.19}$\\ \myrowcolour
GW190503\_185404 & 30.2$_{-4.2}^{+4.2}$ & 0.65$_{-0.23}^{+0.29}$ & -0.03$_{-0.26}^{+0.2}$ & 0.38$_{-0.29}^{+0.42}$ & 0.27$_{-0.11}^{+0.11}$\\
GW190512\_180714 & 14.6$_{-1.0}^{+1.3}$ & 0.54$_{-0.18}^{+0.37}$ & 0.03$_{-0.13}^{+0.12}$ & 0.22$_{-0.17}^{+0.37}$ & 0.27$_{-0.1}^{+0.09}$\\ \myrowcolour
GW190513\_205428 & 21.6$_{-1.9}^{+3.8}$ & 0.5$_{-0.18}^{+0.42}$ & 0.11$_{-0.17}^{+0.28}$ & 0.31$_{-0.23}^{+0.39}$ & 0.37$_{-0.13}^{+0.13}$\\
GW190517\_055101 & 26.6$_{-4.0}^{+4.0}$ & 0.68$_{-0.29}^{+0.27}$ & 0.52$_{-0.19}^{+0.19}$ & 0.49$_{-0.29}^{+0.3}$ & 0.34$_{-0.14}^{+0.24}$\\ \myrowcolour
GW190519\_153544 & 44.5$_{-7.1}^{+6.4}$ & 0.61$_{-0.19}^{+0.28}$ & 0.31$_{-0.22}^{+0.2}$ & 0.44$_{-0.29}^{+0.35}$ & 0.44$_{-0.14}^{+0.25}$\\
GW190521\_074359 & 32.1$_{-2.5}^{+3.2}$ & 0.78$_{-0.21}^{+0.19}$ & 0.09$_{-0.13}^{+0.1}$ & 0.4$_{-0.29}^{+0.32}$ & 0.24$_{-0.1}^{+0.07}$\\\myrowcolour
GW190602\_175927 & 49.1$_{-8.5}^{+9.1}$ & 0.71$_{-0.33}^{+0.25}$ & 0.07$_{-0.24}^{+0.25}$ & 0.42$_{-0.31}^{+0.41}$ & 0.47$_{-0.17}^{+0.25}$\\ 
GW190620\_030421 & 38.3$_{-6.5}^{+8.3}$ & 0.62$_{-0.27}^{+0.32}$ & 0.33$_{-0.25}^{+0.22}$ & 0.43$_{-0.28}^{+0.36}$ & 0.49$_{-0.2}^{+0.23}$\\ \myrowcolour
GW190630\_185205 & 24.9$_{-2.1}^{+2.1}$ & 0.68$_{-0.22}^{+0.27}$ & 0.1$_{-0.13}^{+0.12}$ & 0.32$_{-0.23}^{+0.31}$ & 0.18$_{-0.07}^{+0.1}$\\
GW190701\_203306 & 40.3$_{-4.9}^{+5.4}$ & 0.76$_{-0.31}^{+0.21}$ & -0.07$_{-0.29}^{+0.23}$ & 0.42$_{-0.31}^{+0.42}$ & 0.37$_{-0.12}^{+0.11}$\\ \myrowcolour
GW190706\_222641 & 42.7$_{-7.0}^{+10.0}$ & 0.58$_{-0.25}^{+0.34}$ & 0.28$_{-0.29}^{+0.26}$ & 0.38$_{-0.28}^{+0.39}$ & 0.71$_{-0.27}^{+0.32}$\\ 
GW190707\_093326 & 8.5$_{-0.5}^{+0.6}$ & 0.73$_{-0.27}^{+0.24}$ & -0.05$_{-0.08}^{+0.1}$ & 0.29$_{-0.23}^{+0.39}$ & 0.16$_{-0.07}^{+0.07}$\\ \myrowcolour
GW190708\_232457 & 13.2$_{-0.6}^{+0.9}$ & 0.76$_{-0.28}^{+0.21}$ & 0.02$_{-0.08}^{+0.1}$ & 0.29$_{-0.23}^{+0.43}$ & 0.18$_{-0.07}^{+0.06}$\\ 
GW190720\_000836 & 8.9$_{-0.8}^{+0.5}$ & 0.58$_{-0.3}^{+0.36}$ & 0.18$_{-0.12}^{+0.14}$ & 0.33$_{-0.22}^{+0.43}$ & 0.16$_{-0.06}^{+0.12}$\\ \myrowcolour
GW190727\_060333 & 28.6$_{-3.7}^{+5.3}$ & 0.8$_{-0.32}^{+0.18}$ & 0.11$_{-0.25}^{+0.26}$ & 0.47$_{-0.36}^{+0.41}$ & 0.55$_{-0.22}^{+0.21}$\\ 
GW190728\_064510 & 8.6$_{-0.3}^{+0.5}$ & 0.66$_{-0.37}^{+0.3}$ & 0.12$_{-0.07}^{+0.2}$ & 0.29$_{-0.2}^{+0.37}$ & 0.18$_{-0.07}^{+0.05}$\\\myrowcolour
GW190803\_022701 & 27.3$_{-4.1}^{+5.7}$ & 0.75$_{-0.31}^{+0.22}$ & -0.03$_{-0.27}^{+0.24}$ & 0.44$_{-0.33}^{+0.42}$ & 0.55$_{-0.24}^{+0.26}$\\ 
GW190828\_063405 & 25.0$_{-2.1}^{+3.4}$ & 0.82$_{-0.22}^{+0.15}$ & 0.19$_{-0.16}^{+0.15}$ & 0.43$_{-0.3}^{+0.36}$ & 0.38$_{-0.15}^{+0.1}$\\ \myrowcolour
GW190828\_065509 & 13.3$_{-1.0}^{+1.2}$ & 0.43$_{-0.16}^{+0.38}$ & 0.08$_{-0.16}^{+0.16}$ & 0.3$_{-0.23}^{+0.38}$ & 0.3$_{-0.1}^{+0.1}$\\ 
GW190910\_112807 & 34.3$_{-4.1}^{+4.1}$ & 0.82$_{-0.23}^{+0.15}$ & 0.02$_{-0.18}^{+0.18}$ & 0.4$_{-0.32}^{+0.39}$ & 0.28$_{-0.1}^{+0.16}$\\ \myrowcolour
GW190915\_235702 & 25.3$_{-2.7}^{+3.2}$ & 0.69$_{-0.27}^{+0.27}$ & 0.02$_{-0.25}^{+0.2}$ & 0.55$_{-0.39}^{+0.36}$ & 0.3$_{-0.1}^{+0.11}$\\ 
GW190924\_021846 & 5.8$_{-0.2}^{+0.2}$ & 0.57$_{-0.37}^{+0.36}$ & 0.03$_{-0.09}^{+0.3}$ & 0.24$_{-0.18}^{+0.4}$ & 0.12$_{-0.04}^{+0.04}$\\ \myrowcolour
GW190925\_232845 & 15.8$_{-1.0}^{+1.1}$ & 0.73$_{-0.34}^{+0.24}$ & 0.11$_{-0.14}^{+0.17}$ & 0.39$_{-0.29}^{+0.43}$ & 0.19$_{-0.07}^{+0.07}$\\ 
GW190930\_133541 & 8.5$_{-0.5}^{+0.5}$ & 0.64$_{-0.45}^{+0.3}$ & 0.14$_{-0.15}^{+0.31}$ & 0.34$_{-0.24}^{+0.4}$ & 0.15$_{-0.06}^{+0.06}$\\ \myrowcolour
GW191105\_143521 & 7.8$_{-0.4}^{+0.6}$ & 0.72$_{-0.31}^{+0.24}$ & -0.02$_{-0.09}^{+0.13}$ & 0.3$_{-0.24}^{+0.45}$ & 0.23$_{-0.09}^{+0.07}$\\ 
GW191109\_010717 & 47.5$_{-7.5}^{+9.6}$ & 0.73$_{-0.24}^{+0.21}$ & -0.29$_{-0.31}^{+0.42}$ & 0.63$_{-0.37}^{+0.29}$ & 0.25$_{-0.12}^{+0.18}$\\ \myrowcolour
GW191129\_134029 & 7.3$_{-0.3}^{+0.4}$ & 0.63$_{-0.29}^{+0.31}$ & 0.06$_{-0.08}^{+0.16}$ & 0.26$_{-0.19}^{+0.36}$ & 0.16$_{-0.06}^{+0.05}$\\
GW191204\_171526 & 8.6$_{-0.3}^{+0.4}$ & 0.69$_{-0.26}^{+0.25}$ & 0.16$_{-0.05}^{+0.08}$ & 0.39$_{-0.26}^{+0.35}$ & 0.13$_{-0.05}^{+0.04}$\\ \myrowcolour
GW191215\_223052 & 18.4$_{-1.7}^{+2.2}$ & 0.73$_{-0.27}^{+0.24}$ & -0.04$_{-0.21}^{+0.17}$ & 0.5$_{-0.38}^{+0.37}$ & 0.35$_{-0.14}^{+0.13}$\\ 
GW191216\_213338 & 8.3$_{-0.2}^{+0.2}$ & 0.64$_{-0.29}^{+0.31}$ & 0.11$_{-0.06}^{+0.13}$ & 0.23$_{-0.16}^{+0.35}$ & 0.07$_{-0.03}^{+0.02}$\\ \myrowcolour
GW191222\_033537 & 33.8$_{-5.0}^{+7.1}$ & 0.79$_{-0.32}^{+0.18}$ & -0.04$_{-0.25}^{+0.2}$ & 0.41$_{-0.32}^{+0.41}$ & 0.51$_{-0.26}^{+0.23}$\\ 
GW191230\_180458 & 36.5$_{-5.6}^{+8.2}$ & 0.77$_{-0.34}^{+0.2}$ & -0.05$_{-0.31}^{+0.26}$ & 0.52$_{-0.39}^{+0.38}$ & 0.69$_{-0.27}^{+0.26}$\\ \myrowcolour
GW200112\_155838 & 27.4$_{-2.1}^{+2.6}$ & 0.81$_{-0.26}^{+0.17}$ & 0.06$_{-0.15}^{+0.15}$ & 0.39$_{-0.3}^{+0.39}$ & 0.24$_{-0.08}^{+0.07}$\\ 
GW200128\_022011 & 32.0$_{-5.5}^{+7.5}$ & 0.8$_{-0.3}^{+0.18}$ & 0.12$_{-0.25}^{+0.24}$ & 0.57$_{-0.4}^{+0.34}$ & 0.56$_{-0.28}^{+0.28}$\\ \myrowcolour
GW200129\_065458 & 27.2$_{-2.3}^{+2.1}$ & 0.85$_{-0.41}^{+0.12}$ & 0.11$_{-0.16}^{+0.11}$ & 0.52$_{-0.37}^{+0.42}$ & 0.18$_{-0.07}^{+0.05}$\\ 
GW200202\_154313 & 7.5$_{-0.2}^{+0.2}$ & 0.72$_{-0.31}^{+0.24}$ & 0.04$_{-0.06}^{+0.13}$ & 0.28$_{-0.22}^{+0.4}$ & 0.09$_{-0.03}^{+0.03}$\\ \myrowcolour
GW200208\_130117 & 27.7$_{-3.1}^{+3.6}$ & 0.73$_{-0.29}^{+0.23}$ & -0.07$_{-0.27}^{+0.22}$ & 0.38$_{-0.29}^{+0.41}$ & 0.4$_{-0.14}^{+0.15}$\\ 
GW200209\_085452 & 26.7$_{-4.2}^{+6.0}$ & 0.78$_{-0.31}^{+0.19}$ & -0.12$_{-0.3}^{+0.24}$ & 0.51$_{-0.37}^{+0.39}$ & 0.57$_{-0.26}^{+0.25}$\\ \myrowcolour
GW200219\_094415 & 27.6$_{-3.8}^{+5.6}$ & 0.77$_{-0.32}^{+0.21}$ & -0.08$_{-0.29}^{+0.23}$ & 0.48$_{-0.35}^{+0.4}$ & 0.57$_{-0.22}^{+0.22}$\\ 
GW200224\_222234 & 31.1$_{-2.6}^{+3.2}$ & 0.82$_{-0.26}^{+0.16}$ & 0.1$_{-0.15}^{+0.15}$ & 0.49$_{-0.36}^{+0.37}$ & 0.32$_{-0.11}^{+0.08}$\\ \myrowcolour
GW200225\_060421 & 14.2$_{-1.4}^{+1.5}$ & 0.73$_{-0.28}^{+0.23}$ & -0.12$_{-0.28}^{+0.17}$ & 0.53$_{-0.38}^{+0.34}$ & 0.22$_{-0.1}^{+0.09}$\\ 
GW200302\_015811 & 23.4$_{-3.0}^{+4.7}$ & 0.53$_{-0.2}^{+0.36}$ & 0.01$_{-0.26}^{+0.25}$ & 0.37$_{-0.28}^{+0.45}$ & 0.28$_{-0.12}^{+0.16}$\\ \myrowcolour
GW200311\_115853 & 26.6$_{-2.0}^{+2.4}$ & 0.82$_{-0.27}^{+0.16}$ & -0.02$_{-0.2}^{+0.16}$ & 0.45$_{-0.35}^{+0.4}$ & 0.23$_{-0.07}^{+0.05}$\\ 
GW200316\_215756 & 8.8$_{-0.6}^{+0.6}$ & 0.6$_{-0.38}^{+0.34}$ & 0.13$_{-0.1}^{+0.27}$ & 0.29$_{-0.2}^{+0.38}$ & 0.22$_{-0.08}^{+0.08}$\\ 
 \hline
\end{tabular}
}
\end{table*}


\bsp	
\label{lastpage}
\end{document}